\documentclass{Interspeech2024}

\interspeechcameraready 

\usepackage{
    amsfonts,
    amsmath,
    amssymb,
    colortbl,
    color,
    commath,
    graphicx,
    hyperref,
    wasysym,
    multicol,
    multirow,
    soul,
    times,
    tipa,
    url
}

\definecolor{tablecolor}{cmyk}{0,0,0,0.12}

\title{Fine-Grained and Interpretable Neural Speech Editing}

\name{Max}{Morrison}
\name{Cameron}{Churchwell}
\name{Nathan}{Pruyne}
\name{Bryan}{Pardo}

\address{Northwestern University, US}
\email{morrimax@u.northwestern.edu, pardo@northwestern.edu}


\begin{document}

\maketitle


\begin{abstract}
Fine-grained editing of speech attributes---such as prosody (i.e., the pitch, loudness, and phoneme durations), pronunciation, speaker identity, and formants---is useful for fine-tuning and fixing imperfections in human and AI-generated speech recordings for creation of podcasts, film dialogue, and video game dialogue. Existing speech synthesis systems use representations that entangle two or more of these attributes, prohibiting their use in fine-grained, disentangled editing. In this paper, we demonstrate the first disentangled and interpretable representation of speech with comparable subjective and objective vocoding reconstruction accuracy to Mel spectrograms. Our interpretable representation, combined with our proposed data augmentation method, enables training an existing neural vocoder to perform fast, accurate, and high-quality editing of pitch, duration, volume, timbral correlates of volume, pronunciation, speaker identity, and spectral balance.
\end{abstract}


\section{Introduction}
\label{sec:intro}

Deep generative modeling of speech is a mainstay content creation tool for podcasts, audiobooks, film dialogue, and video game dialogue. These applications require speech that accurately expresses the emotion, emphasis, and cadence of the narrative context, as well as accents and dialects appropriate for the cultural context. Speech generation and editing for these applications should permit high-quality, fine-grained, independent control over the relevant attributes (e.g., prosody and pronunciation) by a human user or machine learning model. For human control, these parameters should be interpretable and intuitive.  

If one also solves the inverse problem of producing these interpretable, disentangled control parameters from existing recorded speech, then editing speech becomes an analysis-modification-synthesis process: speech is encoded in the interpretable, disentangled representation; the user modifies the representation; and the modified speech is resynthesized. In this work, we advance an interpretable representation of speech that can be inferred from speech recordings and used with an off-the-shelf speech synthesis model to create a versatile speech editor amenable to post-production for podcasts, film dialogue, and more. We first overview existing representations for speech generation that precede our proposed representation.

\begin{figure}[t]
    \centering
    \includegraphics[width=\linewidth]{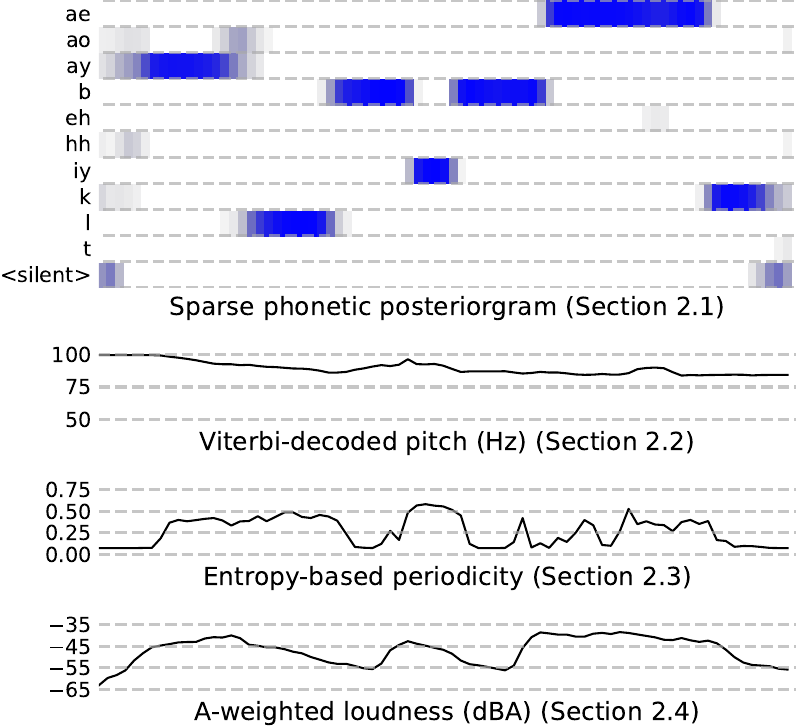}
    \caption{\textbf{Our proposed speech representation $|$} The time-varying components of our interpretable, disentangled speech representation applied to a recording of Arnold Schwarzenegger saying ``I'll be back'' from the movie \textnormal{The Terminator}.}
    \label{fig:representation}
    \vspace{-1.8em}
\end{figure}
\noindent
\textbf{Time-frequency representations $|$} Time-varying representations of frequencies (e.g., a Mel spectrogram) are widely-used and interpretable speech representations that have been used in loss functions~\cite{kong2020hifi}, as output representations~\cite{shen2018natural}, and input representations~\cite{lorenzo2018towards,mustafa2021stylemelgan}. However, time-frequency representations are not disentangled: there is no simple way to edit them to independently change, e.g., the pronunciation or pitch. Wang et al.~\cite{wang2022controllable} propose to disentangle pitch, energy, and speaker from the Mel spectrogram by tuning the number of bottleneck channels to remove information. We omit the bottleneck and replace the Mel spectrogram with an interpretable pronunciation representation that enables fine-grained pronunciation editing~\cite{churchwell2024high}.

Morise et al.~\cite{morise2016world} propose analysis-modification-synthesis of speech using (1) a time-frequency representation, (2) aperiodicity, and (3) pitch. Time-Domain Pitch-Synchronous Overlap-and-Add (TD-PSOLA)~\cite{moulines1990pitch} is a non-parametric method that first segments the audio at the start of each repetition in the waveform (or at equal intervals in unvoiced regions) and uses overlap-add to combine modified (e.g., repeated) audio frames. Both WORLD and TD-PSOLA are digital signal processing (DSP) methods that modify formants (i.e., the relative energy at harmonics in voiced frames) in undesirable ways.

\noindent
\textbf{Lexical representations $|$} Lexical representations such as graphemes (characters) and phonemes (discrete units of speech sound) are inputs for text-to-speech (TTS) systems~\cite{arik2017deep}. Ren et al.~\cite{ren2020fastspeech} demonstrate TTS with phoneme duration control; Łańcucki~\cite{lancucki2021fastpitch} demonstrates pitch control. However, when used for analysis-modification-synthesis, graphemes and phonemes induce coarse discretization, causing ambiguous pronunciation. Further, TTS phoneme inputs are often produced from text, ignoring the input speech pronunciation and requiring a transcript.

\noindent
\textbf{Latent representations $|$} Latent speech representations are non-interpretable and typically not disentangled. These representations exhibit strong performance in speech reconstruction~\cite{kumar2023high} and generation~\cite{wang2023neural}. However, precise control of, e.g., pitch is difficult, as pitch is entangled within a non-interpretable representation. Notable exceptions are representations with partial speaker disentanglement~\cite{kong2023encoding}, pitch-agnostic latents of automatic speech recognition (ASR) models~\cite{choi2022nansy++,kovela2023any}, and discrete factorizations~\cite{ju2024naturalspeech}. However, lack of interpretable pronunciation representation prohibits fine-grained pronunciation control.

\noindent
\textbf{Source-filter representations $|$} Neural source filter (NSF) 
methods~\cite{wang2019neural, morrison2022neural, yoneyama2023source, ohtani2024firnet} represent speech as a periodic source excitation and a time-varying FIR filter. NSF methods demonstrate fast, accurate, and high-fidelity pitch-shifting. However, the speed of analysis-modification-synthesis with existing NSF models is constrained by slow pitch estimators and causal operations at the waveform resolution. As well, no NSF model has demonstrated pronunciation or spectral balance editing. No model using any representation has demonstrated disentangled control of volume from the timbral correlates of volume.

\noindent
\textbf{Our primary contributions are as follows.}
\begin{itemize}
    \item \textbf{(Contribution 1)} We are the first to show a transcript-free speech representation (Section~\ref{sec:representation}) that is interpretable, disentangled, and amenable to accurate speech reconstruction (Section~\ref{sec:reconstruction}) and high-quality editing (Section~\ref{sec:eval-entangle}).
    \item \textbf{(Contribution 2)} We introduce a novel data augmentation method (Section~\ref{sec:augment}) that improves speech quality, enables spectral balance control, allows pitch-shifting outside the range of training data, and disentangles volume from the timbral correlates of volume.
    \item \textbf{(Contribution 3)} We develop a single model capable of fast, accurate, high-fidelity editing of pitch, duration, volume, timbral correlates of volume, pronunciation, speaker, and spectral balance.
\end{itemize}
We provide audio examples and open-source code on our project website.\footnote{\url{maxrmorrison.com/sites/promonet}}


\section{A disentangled, interpretable representation of speech}
\label{sec:representation}

Our speech representation (Figure~\ref{fig:representation}) can be computed directly from a speech recording (without need of a transcript) and consists of four disentangled, interpretable, time-aligned features: sparse phonetic posteriorgrams (SPPG) (Section~\ref{sec:ppg}), Viterbi-decoded pitch (Section~\ref{sec:pitch}), entropy-based periodicity (Section~\ref{sec:periodicity}), and multi-band A-weighted loudness (Section~\ref{sec:loudness}).


\subsection{Sparse phonetic posteriorgrams (SPPGs)}
\label{sec:ppg}

A phonetic posteriorgram (PPG) is a time-varying distribution over acoustic units of speech (e.g., phones or phonemes)~\cite{hazen2009query}. We infer PPGs over the 40 phonemes of the CMU pronunciation dictionary\footnote{\url{speech.cs.cmu.edu/cgi-bin/cmudict}} from Mel spectrograms. These PPGs allow good independent control of pitch and pronunciation~\cite{churchwell2024high}.

We hypothesize that noisy regions in our representation (e.g., low-probability phoneme bins in the PPG) can be memorized, inducing overfitting that harms generalization during reconstruction and editing. We address this by proposing \textit{sparse phonetic posteriorgrams} (SPPGs). We explore three methods for producing SPPGs: (1) (\textbf{Top-$k$}) set all but the $k$ most-probable phonemes at each frame to zero; (2) (\textbf{Threshold-$k$}) set all phonemes with probability less than $k$ to zero; and (3) (\textbf{Percentile-$k$}) sort the phonemes of each frame by descending probability, add phoneme probabilities until the sum reaches $k$, and set all remaining phoneme probabilities to zero. We renormalize SPPG frames to sum to one. Our hyperparameter search over $k$ for each method indicated Percentile-$k$ with $k=0.85$ is best,  using objective metrics defined in Section~\ref{sec:metrics}.


\subsection{Viterbi-decoded pitch}
\label{sec:pitch}
We compute our pitch representation using the FCNF0++ pitch estimator~\cite{morrison2023cross}. FCNF0++ produces a pitch sequence $y = y_1, \dots, y_T$ in Hz from a time-varying categorical posterior distribution $D \in \mathbb{R}^{|Q| \times T}$ inferred by a neural network over $T$ time frames and a set of frequency values $Q$. FCNF0++ uses $|Q| = 1440$ frequency bins, ranging from 31-2006 Hz in 5 cent intervals. Cents ($\cent$) measures the pitch difference between two frequency values $y_t$ and $\hat{y}_t$: $\cent\left(y_t, \hat{y}_t\right) = 1200 \left| \log_2 \left(y_t / \hat{y}_t\right)\right|$. By default, FCNF0++ decodes $y$ from time-varying distribution $D$ using argmax in voiced frames $\mathcal{V}$ and linear interpolation of the closest voiced pitch values within unvoiced frames. Voiced frames are time frames where the periodicity $h = h_1, \dots, h_T$ (Section~\ref{sec:periodicity}) exceeds a threshold: $\mathcal{V} = \{t: h_t > \alpha\}$. We find there is no threshold $\alpha$ sufficient to remove the noise in unvoiced frames while leaving voiced frames unmodified. We use Viterbi decoding~\cite{viterbi1967error}, which finds the optimal path of a time-varying distribution given initial and transition probabilities. Our initial probabilities are uniform and our transition probabilities are triangular distributions that assign maximal probability to staying on the same pitch and zero probability to pitch jumps greater than one octave between adjacent time frames. We could not find an open-source Viterbi decoder fast enough to scale to large datasets. We develop and open-source\footnote{\texttt{\href{https://github.com/maxrmorrison/torbi}{github.com/maxrmorrison/torbi}}}
a Viterbi decoder that decodes time-varying distribution $D$ on VCTK~\cite{yamagishi2019cstr} 1.62x faster than a widely-used reference~\cite{mcfee2015librosa} on a 16-core CPU. Using a batch size of 1, our GPU implementation on an A40 GPU is 1,760x faster than our 16-core CPU implementation. Using a batch size of 512, our GPU implementation is 309,000x faster than a 16-core CPU (501,000x faster than reference).

Pitch conditioning of neural networks often involves quantizing pitch into equal-width bins and learning a corresponding embedding table~\cite{valin2019lpcnet, morrison2022neural, wang2022controllable}. This leads to infrequently used bins at endpoints of the pitch range of the training distribution that cause instability and artifacts. We change the bin spacing so each pitch bin is accessed equally often during training. This produces a variable-width quantization that allocates more bins to frequently used pitch regions (e.g., 100 to 200 Hz). We use 256 bins and a 64-dimensional embedding table.



\subsection{Entropy-based periodicity}
\label{sec:periodicity}

We use entropy-based periodicity~\cite{morrison2023cross} estimation. Given audio frame $x_t$, its periodicity $h_t \in [0, 1]$ indicates the extent to which $x_t$ contains pitch (i.e., is \textit{voiced}): $h_t = 1 - \frac{1}{\ln |Q|} \sum_{q \in Q}D_{q, t} \ln D_{q, t}$. Higher values indicate that $x_t$ is likely to contain pitch. Unlike binary voiced/unvoiced (V/UV) masks, periodicity encodes the uncertainty of V/UV decisions.


\subsection{Multi-band A-weighted loudness}
\label{sec:loudness}

A-weighted loudness~\cite{mccurdy1936tentative} is a frequency-average of a weighted magnitude spectrogram, with per-channel weights derived from human perceptual studies of loudness variation. Reconstructing speech from our representation using A-weighted loudness produces worse loudness reconstruction relative to Mel spectrograms (Table~\ref{tab:results}; w/o multi-band), but Mel spectrogram vocoding entangles pitch and pronunciation. To address this tradeoff, we propose using \textit{multi-band A-weighted loudness}: sorted, real-valued FFT frequencies $\omega_1, \dots, \omega_{|\omega| / 2 + 1}$ are divided into $k$ bands and averaging occurs over each band. A hyperparameter search over 2, 4, 8, 16, and 32 bands indicates optimal disentanglement and loudness reconstruction at 8 bands.


\section{Controlling spectral balance and timbre}
\label{sec:augment}

Prior works use resampling~\cite{morrison2020controllable} or pitch-shifting~\cite{bae2022enhancement} to augment the pitch distribution of a training dataset of speech. However, these methods cannot be used in speaker-conditioned generation without causing artifacts such as incorrect formants. We propose a technique that increases the range of the desired audio feature within the training distribution and enables the artifacts induced by data augmentation to be independently controlled. We apply our proposed augmentation technique to disentangle pitch (F0) from spectral balance, as well as the disentanglement of volume from the timbral correlates of volume.

\noindent
\textbf{Disentangling spectral balance and pitch} $|$  Let $R_f(x; a, b)$ resample speech recording $x$ from sampling rate $a$ to sampling rate $b$. Given original sampling rate $s$, target sampling rate $t$, and random pitch shift factor $r_f \sim \text{Uniform}(-1, 1)$, we augment training data with $x_f = R_f(R_f(x; 2^{r_f}s, s); s, t)$. This augmentation modifies spectral balance. We pass $r_f$ to the network during training, so the model learns that low $r_f$ indicates more energy at low-frequencies and high $r_f$ indicates more energy at high-frequencies. We create one randomly pitch-shifted copy of each training utterance.

\noindent
\textbf{Disentangling volume from its timbral correlates} $|$ Let $R_l(x; g)$ be a function that increases or decreases the volume of speech recording $x$ by $g$ decibels. Given a randomly sampled volume shift $r_l \sim \text{Uniform}(-1, 1)$, we augment training data with $x_l = R_l(x; 12 r_l)$. If any sample of $x_l$ is outside $[-1, 1]$, we draw a new sample for $r_l$ until $x_l$ is within $[-1, 1]$. We pass $r_l$ to the network during training. We create one randomly volume-shifted copy of each training utterance. During generation, setting $r_l > 0$ increases volume, $r_l < 0$ decreases volume, and $r_l = 0$ maintains the current volume; $r_l$ does not control the timbral correlates. Instead, framewise edits to A-weighted loudness produce audible changes in timbre corresponding to louder or quieter speech while maintaining accurate volume control.


\section{Neural speech editing model}
\label{sec:model}

We create a high-quality, fine-grained speech editing model by training an off-the-shelf HiFi-GAN vocoder~\cite{kong2020hifi} on our proposed distentangled, interpretable representation (Section~\ref{sec:representation}), as well as three time-invariant features: augmentation ratios $r_f$ and $r_l$ (Section~\ref{sec:augment}) and a jointly trained speaker embedding. We replace the multi-scale spectrogram discriminator (MSD)~\cite{kong2020hifi} with the complex, multi-band spectrogram discriminator~\cite{kumar2023high}, which allows the discriminator to evaluate phase.

We train for 400k steps on one A40 GPU. We use a batch size of 64. Each item in the batch consists of 64 frames and produces 16,384 samples of synthesized audio. We use the AdamW optimizer~\cite{loshchilov2017decoupled} with a learning rate of $2 \times 10^{-4}$.


\subsection{Data}
\label{sec:data}

We use VCTK~\cite{yamagishi2019cstr} for training and evaluation. We select five male and five female test speakers. We select ten utterances from each test speaker for a total of 100 test utterances. We require all test utterances be between four and ten seconds in length to provide adequate context for perceptual studies. We reserve 64 random validation utterances. On our companion website, we show our model is also capable of adaptation to unseen speakers from the DAPS~\cite{mysore2014can} dataset.


\begin{table*}[ht]
\centering
\begin{tabular}{c|l|cccccl}
\textbf{Task} & \textbf{Method} & $\boldsymbol{\Delta\cent}\downarrow$ & $\boldsymbol{\Delta\phi}\downarrow$ & $\boldsymbol{\Delta}$\textbf{dBA}$\downarrow$ & $\boldsymbol{\Delta}$\textbf{PPG}$\downarrow$ & \textbf{Subjective}$\uparrow$\\
\rowcolor{tablecolor}
& \textbf{Proposed} & \textbf{17.1} $\boldsymbol{\pm}$ \textbf{0.88} & \textbf{.055} $\boldsymbol{\pm}$ \textbf{.003} & .521 $\pm$ .039 & .109 $\pm$ .008 & .471 $\pm$ .046\\
\rowcolor{tablecolor}
\multirow{-2}{*}{\textbf{Reconstruct}}& Mels & 21.5 $\pm$ 1.44 & .061 $\pm$ .003 & \textbf{.381} $\boldsymbol{\pm}$ \textbf{.019} & \textbf{.041} $\boldsymbol{\pm}$ \textbf{.003} & .529 $\pm$ .046 \\

& \textbf{Proposed} & 22.5 $\pm$ 1.41 & \textbf{.082} $\boldsymbol{\pm}$ .003 & \textbf{.874} $\boldsymbol{\pm}$ .066  & .130 $\pm$ .007 &  \textbf{68.9} \\
& TD-PSOLA~\cite{moulines1990pitch} & 22.4 $\pm$ 1.35 & .112 $\pm$ .005 & 1.32 $\pm$ .066 & \textbf{.109} $\boldsymbol{\pm}$ \textbf{.006} & 61.1 $\pm$ 3.13 \\
\multirow{-3}{*}{\textbf{Pitch-shift}} & WORLD~\cite{morise2016world} & \textbf{18.6} $\boldsymbol{\pm}$ \textbf{0.72} & .113 $\pm$ .005 & 1.62 $\pm$ .056 & .296 $\pm$ .045 & 45.0 $\pm$ 3.25 \\

\rowcolor{tablecolor}
& \textbf{Proposed} & \textbf{19.7} $\boldsymbol{\pm}$ \textbf{0.74} & \textbf{.068} $\boldsymbol{\pm}$ \textbf{.002} & .720 $\pm$ .030 & \textbf{.142} $\boldsymbol{\pm}$ \textbf{.005} & \qquad -- \\
\rowcolor{tablecolor}
& \enskip w/o SPPG & 20.0 $\pm$ 0.62 & .070 $\pm$ .002 & .754 $\pm$ .030 & .146 $\pm$ .005 & \qquad -- \\
\rowcolor{tablecolor}
& \enskip w/o Viterbi decoding & 34.2 $\pm$ 1.57 & .071 $\pm$ .002 & .709 $\pm$ .028 & .143 $\pm$ .005 &  \qquad -- \\
\rowcolor{tablecolor}
& \enskip w/o variable-width bins & 20.5 $\pm$ 0.69 & \textbf{.068} $\boldsymbol{\pm}$ \textbf{.002} & .728 $\pm$ .031 & .143 $\pm$ .005 & \qquad -- \\
\rowcolor{tablecolor}
& \enskip w/o multi-band loudness & 20.5 $\pm$ 0.64 & .071 $\pm$ .002 & .811 $\pm$ .025 & .166 $\pm$ .005 & \qquad -- \\
\rowcolor{tablecolor}
& \enskip w/o augmentation & 20.2 $\pm$ 0.82 & .070 $\pm$ .002 & \textbf{.679} $\boldsymbol{\pm}$ \textbf{.025} & \textbf{.142} $\boldsymbol{\pm}$ \textbf{.005} & \qquad -- \\
\rowcolor{tablecolor}
\multirow{-7}{*}{\textbf{Ablations}}& \enskip w/o all (cumulative) & 37.5 $\pm$ 1.65 & .071 $\pm$ .002 & .856 $\pm$ .030 & .166 $\pm$ .006 & \qquad -- \\
\end{tabular}
\caption{\textbf{Evaluation results $|$} \textbf{(Reconstruct; Section~\ref{sec:reconstruction})} Results of speech reconstruction using Mel spectrograms or our disentangled, interpretable representation. \textbf{(Pitch-shift; Section~\ref{sec:eval-entangle})} Results of pitch-shifting by $\pm 600$ cents using our proposed system and two DSP baselines. 
\textbf{(Ablations)} Non-cumulative ablations of methods proposed in Sections~\ref{sec:representation} and~\ref{sec:augment}; values are averages over pitch-shifting (by $\pm600$ cents), time-stretching (by factors $\sqrt{2}$ and $\sqrt{2}/2$), loudness edits (by $\pm$5 dBA), and reconstruction. 
}
\label{tab:results}
\vspace{-2.2em}
\end{table*}

\section{Evaluation}
\label{sec:evaluation}

We design our evaluation to answer three questions: (1) Does our representation permit vocoding with comparable objective and subjective quality to Mel spectrograms? (Section~\ref{sec:reconstruction}), (2) Does our representation enable accurate, high-quality prosody control? (Section~\ref{sec:eval-entangle}) and (3) Does our data augmentation method (Section~\ref{sec:augment}) enable disentanglement of pitch and spectral balance, as well as disentanglement of volume from its timbral correlates? (Section~\ref{sec:eval-augment}) Audio examples for all evaluations as well as voice conversion and fine-grained pronunciation editing are available on our project website.


\subsection{Objective metrics}
\label{sec:metrics}

We use four objective metrics: (1) pitch error in cents in voiced regions ($\boldsymbol{\Delta\cent}$), (2) periodicity error as RMSE ($\boldsymbol{\Delta\phi}$), (3) volume error in non-silent frames (above -60 dBA) in decibels ($\boldsymbol{\Delta}$\textbf{dBA}), and (4) Jensen-Shannon divergence between sparsified PPGs ($\boldsymbol{\Delta}$\textbf{PPG})~\cite{churchwell2024high}, a pronunciation distance that strongly correlates with word error rate using Whisper~\cite{radford2022whisper}. 


\subsection{Crowdsourced subjective evaluation}
\label{sec:subjective}

For all crowdsourced evaluations, we use Reproducible Subjective Evaluation (ReSEval v0.1.6)~\cite{morrison2022reproducible} to deploy Human Intelligence Tasks (HITs) on Amazon Mechanical Turk (AMT). For each evaluation, we recruit US participants with a minimum 99\% approval rating and 1000 completed assignments. On average, we pay participants \$13.35 per hour. We omit participants who fail a prescreening listening test. We use as source speech the 100 VCTK~\cite{yamagishi2019cstr} test utterances (Section~\ref{sec:data}).


\subsection{Evaluation of speech reconstruction}
\label{sec:reconstruction}

Mel spectrograms excel at speech reconstruction. We compare our disentangled, interpretable representation to Mel spectrograms on speech reconstruction with HiFi-GAN~\cite{kong2020hifi}. We include speaker conditioning, data augmentation, and the complex, multi-band discriminator~\cite{kumar2023high} in our baseline Mel model. This is for fair comparison and because these techniques improve the baseline. We use objective metrics described in Section~\ref{sec:metrics}. For subjective evaluation, we recruit 35 participants to each perform 15 ABX comparisons of reconstruction accuracy. In an ABX comparison, a participant selects which of two speech recordings (``A'' or ``B'') sounds more similar to a reference recording (``X''). All participants passed the listening test and five participants left early, giving us 450 ABX comparisons. Table~\ref{tab:results} (\textbf{Reconstruct}) shows that participants rated our representation more similar to ground truth audio in 212 of 450 ABX comparisons ($47.1\%$). A two-sided Binomial test indicates no significant preference among raters ($p = 0.23$); the perceptual reconstruction accuracy of our representation is roughly as good as Mel spectrograms. As well, our representation improves pitch and periodicity reconstruction relative to Mel spectrograms. We next demonstrate the efficacy of our representation on editing tasks that are challenging or impossible with other representations (e.g., Mel spectrograms).


\subsection{Evaluation of disentangled prosody control}
\label{sec:eval-entangle}
We demonstrate disentanglement of pitch by modifying the pitch by $\pm 600$ cents (i.e., one tritone) while keeping all other features the same. We demonstrate duration control by increasing or decreasing the speaking rate by a factor of $\sqrt{2}$ via interpolation. We use spherical linear interpolation (SLERP)~\cite{shoemake1985animating} to interpolate PPGs (prior to sparsification) and linear interpolation for all other features. As is the case for human speakers, we only apply time-stretching to voiced phonemes and silences. We use the four objective metrics described in Section~\ref{sec:metrics} for pitch-shifting. We also perform subjective evaluations. Participants listen to speech recordings from each condition and rank their relative quality from 0 (worst) to 100 (best) using a slider. We compare three conditions: (1) our best system, (2) TD-PSOLA~\cite{moulines1990pitch} and, (3) WORLD~\cite{morise2016world}. We recruit 35 participants for each evaluation (pitch-shifting and time-stretching). Five participants failed the prescreening listening test for the pitch-shifting and one participant left early, so we receive 435 three-way comparisons for pitch-shifting and 525 for time-stretching. Table~\ref{tab:results} (\textbf{Pitch-shift}) indicates our proposed method demonstrates statistically significant improvements in perceptual pitch-shifting quality using a Wilcoxon signed-rank test ($p=5.41 \times 10^{-6}$). Our \textbf{time-stretching} subjective evaluation indicates insignificant improvements in time-stretching ($p=0.45$), with a mean of 64.0 for our proposed method, 63.3 $\pm$ 1.71 for TD-PSOLA, and 46.5 $\pm$ 2.20 for WORLD.

We demonstrate that our design choices (Sections~\ref{sec:representation} and~\ref{sec:augment}) improve editing accuracy via ablations. We report the average objective metrics over a set of modifications: pitch-shifting, time-stretching, loudness adjustments, and reconstruction. Table~\ref{tab:results} (\textbf{Ablations}) shows that each of our design decisions contributes to the efficacy of our proposed system, with Viterbi-decoded pitch (Section~\ref{sec:pitch}) and multi-band A-weighted loudness (Section~\ref{sec:loudness}) being particularly impactful.


\subsection{Evaluation of data augmentation}
\label{sec:eval-augment}

We now demonstrate that our proposed data augmentation (Section~\ref{sec:augment}) permits disentangling spectral balance from pitch, as well as disentangling volume from its timbral correlates. We perform vocoding using two spectral balance editing ratios: $r_f = \sqrt{2}$ and $r_f = \sqrt{2}/2$. We perform estimation of F0 and its first two harmonics (H1 and H2) and measure the displacement of F0, H1, and H2 reconstruction in cents in voiced regions. We also measure the change in spectral centroid between ground-truth and edited audio. Low displacement error with a change in spectral centroid in the direction of $r_f$ indicates disentangled control of spectral balance. Prior methods such as peak-picking~\cite{broad1989formant}, Viterbi decoding~\cite{kopec1985formant}, or neural methods~\cite{dissen2019formant} exhibit significant noise. No prior work has combined neural networks and Viterbi-based harmonic estimation. We propose using our pitch representation (Section~\ref{sec:pitch}) as F0 and performing Viterbi decoding on the log magnitude spectrogram in bands, where H$i$ is restricted to band $(i + w) \times \text{F}0 < \text{F}i < (i + 1/w) \times \text{F}0$ and $w=4/5$ is tuned by visual inspection on training data to prevent octave errors. Our proposed estimation method reconstructs F0 with an average error of 18.73 cents and H1 and H2 with an average error of 5.60 cents. The change in framewise spectral centroid has Pearson correlation of .853 with $r_f$, indicating strong, disentangled control of spectral balance. Qualitatively, this produces a similar effect as, e.g., Alvin from \textit{Alvin and the Chipmunks}, but without requiring voice actors to sing/speak unnaturally slowly.

For disentangling volume from its timbral correlates, we perform vocoding using A-weighted loudness contours modified by $\pm 10$ dBA. We then use simple gain scaling to perform framewise A-weighted volume matching between our loudness-edited speech recordings and corresponding original speech. This produces pairs of original and loudness-edited speech recordings with equal volume. We perform an A/B subjective evaluation in which we ask 35 participants to select which of two speech recordings ``sounds like the speaker is speaking \textbf{louder}''. Six participants failed the prescreening listening test, so we receive 435 perceptual A/B comparisons. In 290 comparisons ($66.7 \pm 4.6\%$), participants' selection matches our intended modification on volume-matched audio ($p=3.20\times 10^{-12}$).


\section{Conclusion}
\label{sec:conclusion}

We demonstrate an interpretable, disentangled speech representation (Section~\ref{sec:representation}) with reconstruction accuracy comparable to Mel spectrograms (Section~\ref{sec:reconstruction}). Our representation and data augmentation method (Section~\ref{sec:augment}) enable off-the-shelf vocoders to perform accurate, high-quality control over pitch, duration, volume, timbral correlates of volume, pronunciation, speaker, and spectral balance (Sections~\ref{sec:eval-entangle}-\ref{sec:eval-augment}). Future work includes one-shot disentangled voice conversion~\cite{kong2023encoding}; generating our representation from lexical features; and framewise formant control using our proposed harmonic estimation (Section~\ref{sec:eval-augment}).


\bibliographystyle{IEEEtran}
\bibliography{refs}

\end{document}